\documentclass[journal=nalefd,manuscript=letter,layout=twocolumn]{achemso}

\usepackage[version=3]{mhchem} 

\pdfoutput=1

\usepackage[english]{babel}
\usepackage[utf8]{inputenc}
\usepackage{newtxtext, newtxmath}
\usepackage{soul}
\usepackage[scaled=0.8]{beramono}
\usepackage[T1]{fontenc}
\usepackage{breqn}

\usepackage{acronym}
\usepackage{bm}
\usepackage{dsfont}
\usepackage{graphicx}
\usepackage{mathtools}
\usepackage{microtype}
\usepackage{tikz}

\usetikzlibrary{decorations.markings}
\usetikzlibrary{decorations.pathmorphing}

\definecolor{mauve}{RGB}{94, 60, 153}

\usepackage[colorlinks, allcolors=mauve]{hyperref}
\usepackage[figure, figure*]{hypcap}

\def\Zagreb{Centre for Advanced Laser Techniques, Institute of Physics, 10000 Zagreb, Croatia}

\author{Nina Girotto}
\affiliation\Zagreb

\author{Dino Novko}
\email{dino.novko@gmail.com}
\affiliation\Zagreb

\title{Dynamical Phonons Following Electron Relaxation Stages in Photo-excited Graphene}


\begin{document}

\sloppy

\begin{abstract}
Ultrafast electron-phonon relaxation dynamics in graphene hides many distinct phenomena, such as hot phonon generation, dynamical Kohn anomalies, and phonon decoupling, yet still remains largely unexplored. Here, we unravel intricate mechanisms governing the vibrational relaxation and phonon dressing in graphene at a highly non-equilibrium state by means of first-principles techniques. We calculate dynamical phonon spectral functions and momentum-resolved linewidths for various stages of electron relaxation and find photo-induced phonon hardening, overall increase of relaxation rate and nonadiabaticity as well as phonon gain. Namely, the initial stage of photo-excitation is found to be governed by strong phonon anomalies of finite-momentum optical modes along with incoherent phonon production. Population inversion state, on the other hand, allows production of coherent and strongly-coupled phonon modes. Our research provides vital insights into the electron-phonon coupling phenomena in graphene, and serves as a foundation for exploring non-equilibrium phonon dressing in materials where ordered states and phase transitions can be induced by photo-excitation.
\end{abstract}

\vspace{10mm}

Through the ionic motion manipulation, photoexcitation as in pump-probe setup is a powerful tool which paves the way for a highly effective designing and customizing the desired functionalities of materials~\cite{lattice_control,torre21}. 
Namely, it can induce novel phases~\cite{Basov2017}, sometimes unreachable in equilibrium, opening, for example, the possibility of light-induced superconductivity~\cite{fausti11,mitrano16}, charge-density-wave order\,\cite{stojchevska14,kogar20}, ferroelectricity\,\cite{nova19}, and disorder-assisted structural transition\,\cite{VO2}. Very often, these states of matter are characterized with a strongly-coupled phonon mode and considerable electron-phonon coupling (EPC), which are believed to be additionally renormalized in a photo-excited non-equilibrium regime\,\cite{gierz17,shengmeng22}.
For instance, photo-induced softening of the relevant phonon mode is quite common in ultrafast dynamics\,\cite{hase02,bismuth2}, nonetheless, photo-excitation can in some cases lead to phonon hardening and consequently stabilize the structural phase\,\cite{Ishioka2008,yan2009,KTaO3,Jiang2016TheOO,otto21}.

Graphene exhibits extraordinary mechanical, transport and optoelectronic properties~\cite{RevModPhys.86.959}, and is therefore an ideal platform to investigate the fundamentals of ultrafast electron-lattice dynamics. Experimental techniques such as time- and angle-resolved photoemission spectroscopy (tr-ARPES)\,\cite{bauer_18, snapshots,direct_view,na19,reutzel22},  two-photon photoemission~\cite{xu96,moos01,PhysRevX.7.011004}, and transient optical spectroscopy~\cite{scops_graphite,breusing09,pagliara11} have reveled various aspects of carrier thermalization in graphene, such as rapid electron-electron and electron-phonon recombination of highly non-equilibrium distribution towards population inversion\,\cite{snapshots}, scattering of electrons with strongly-coupled optical phonons\,\cite{direct_view,kampfrath05}, cooling of hot carriers via acoustic phonons\,\cite{direct_view}, as well as electron band renormalization\,\cite{pagliara11}. 
On the other hand, non-equilibrium phonon dynamics in graphene, phonon relaxation pathways and the corresponding photo-induced phonon dressing are less investigated.
Raman and coherent phonon spectroscopy have demonstrated considerable phonon hardening of the $E_{2g}$ optical phonon mode\,\cite{Ishioka2008,yan2009,ferrante18}, which was attributed to the reduction of the nonadiabatic electron-phonon interaction in non-equilibrium\,\cite{Ishioka2008}. Recent attosecond core-level spectroscopy also uncovered ultrafast phonon stiffening of both zone-center $E_{2g}$ and zone-edge $A_{1}'$ phonon Kohn anomalies\,\cite{biegert21}.

The theoretical studies of the aforesaid ultrafast phenomena are mostly based on the two- and multi-temperature models\,\cite{direct_view,fabio_novko_20,fabio_novko_22}, as well as on the time-dependent Boltzmann equations\,\cite{breusing09,tong21}, and were proven to be valuable in comprehending energy transfer between electrons and strongly-coupled hot phonons. However, these methods do not account for transient phonon renormalization and as such are not suitable for exploring all aspects of EPC and phonon dynamics far from equilibrium, such as structural transitions and soft phonon physics. The phonon renormalization in graphene was recently inspected by means of real-time time-dependent density functional theory in combination with molecular dynamics and real-space lattice distortions, which allows for time-resolved self-consistent renormalization of phonons and EPC strengths\,\cite{shengmeng22}. However, since it relies on real-space distortions within the supercell approach it is able to track the phonon dynamics of only few selected modes. In addition, the study delivered somehow conflicting results, i.e., instead of phonon hardening and electron-phonon decoupling as observed in the experiments\,\cite{Ishioka2008,yan2009,ferrante18,biegert21}, the phonon softening and enhanced EPC strength were reported\,\cite{shengmeng22}. This calls for further theoretical insights on non-equilibrium phonon dynamics in graphene.

Here, we overcome these difficulties and investigate the effects of the photo-excited population on phonon dynamics and EPC in graphene by means of constrained density functional perturbation theory (cDFPT)\,\cite{Murray2007,Nomura2015}. Important advantage of this approach is that it provides a full momentum-dependent picture of phonon renormalization due to constrained photo-excited electron distribution\,\cite{hbn_cdft,Te_A1_DFT,Murray2007,bismuth3,paillard19,marini_cdfpt}. In addition, we combine cDFPT with nonadiabatic phonon self-energy calculations in order to provide information on phonon relaxation rates (linewidths) and nonadiabatic frequency modifications, which are absent in the standard adiabatic cDFPT studies. We discuss phonon renormalization for the usual stages of carrier relaxation in graphene, namely, for strong non-equilibrium, population inversion, and hot (Fermi-Dirac) carrier distribution. We observe remarkable modifications of the well-known Kohn anomalies at the $\Gamma$ and K points of the Brillouin zone, as well as appearance of the additional phonon anomalies away from the Brillouin zone center induced by non-equilibrium population, renormalizing the equilibrium dispersion for up to 6\,meV. Light-induced increase of the overall phonon linewidth and nonadiabatic effects are observed along with a striking phonon gain, where the latter becomes coherent once graphene reaches the state of photo-inversion. From the fermiology analysis we show that the EPC coupling matrix elements are slightly reduced in non-equilibrium state, while the observed features mostly stem from the modified scattering phase space under transient conditions.  With this work, we expand our microscopic understanding of phonon relaxation dynamics of graphene in far non-equilibrium, which along with the well-explored electron thermalization paths constitutes the full dynamical picture of electron-phonon scatterings in graphene.


\begin{figure}[!t]
\includegraphics[width=0.5\textwidth]{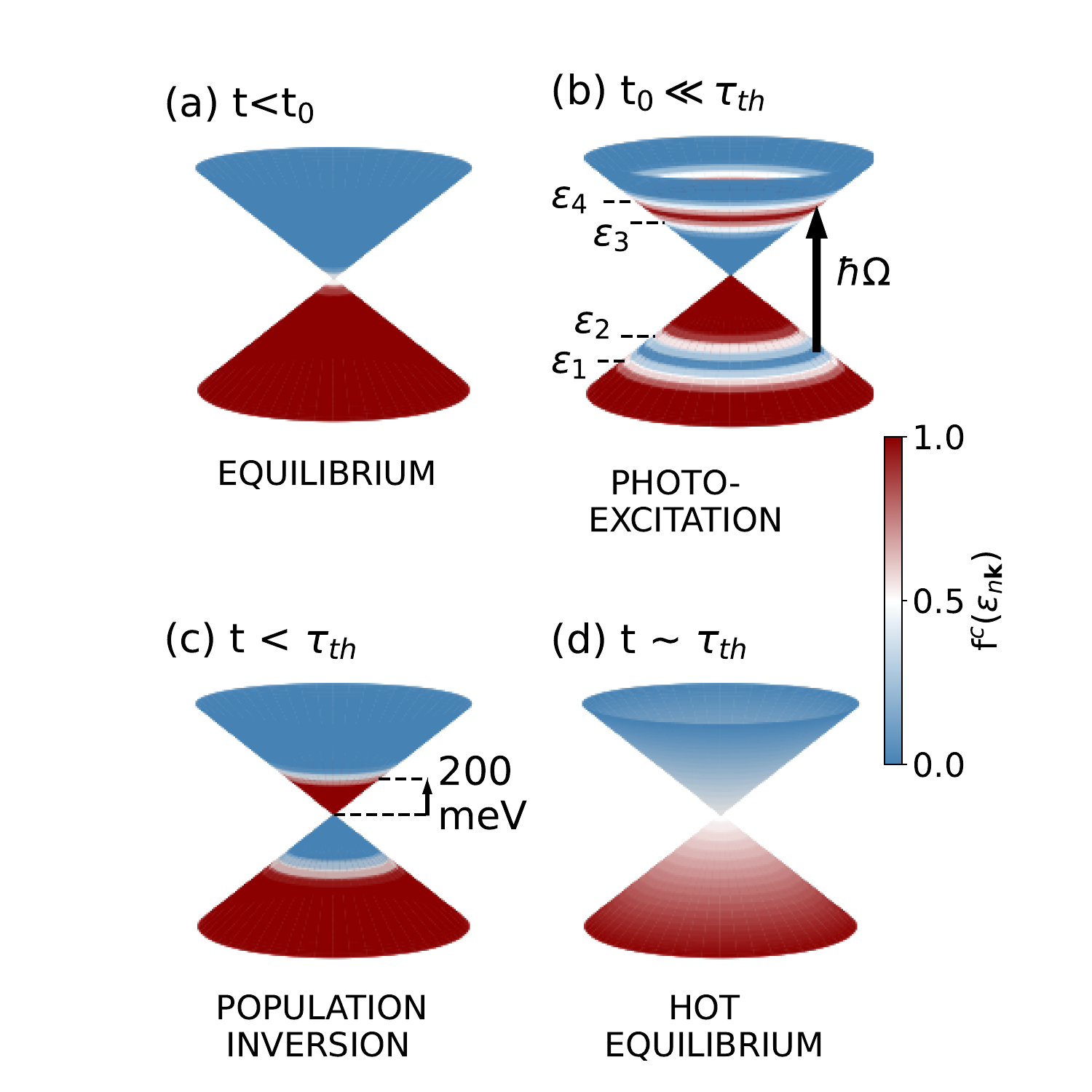}
\caption{Schematic representation of different stages of electron relaxation in photo-excited graphene. (a) Dirac cone with an equilibrium Fermi-Dirac electron distribution. (b) At $t_0$,  laser pulse excites electrons form the [$\varepsilon_1$, $\varepsilon_2$] energy interval vertically upwards in the conduction band, where they fill the states in the [$\varepsilon_3$, $\varepsilon_4$] range. (c) Immediately after the pulse, electrons scatter with other electrons and strongly-coupled phonons ($E_{2g}\simeq200$\,meV and $A'_{1}\simeq 160$\,meV) until the population inversion is created. (d) When the time scale of electron thermalization time ($\tau_{\rm th}$) is reached, electrons follow a hot Fermi-Dirac distribution, which in our calculations, amounts to 2200K. 
} 
\label{fig:fig1}
\end{figure}

Photo-excitation implies promoting a certain carrier density form the valence to the conduction band separated by the energy of laser pulse $\hbar\Omega$, which typically amounts to 1.5\,eV. In this work, we track phonon properties for various electron distributions inside the Dirac cone following directly the pulse application. First, an equilibrium distribution [Fig.~\ref{fig:fig1}\,(a)] is disrupted with a short (fs) pulse causing an occurrence of empty states below the Dirac point and filled states above it [Fig.~\ref{fig:fig1}\,(b)]. Photo-generated electrons and holes establish separate distributions and begin a rapid process of thermalization and cooling through carrier-carrier and carrier-phonon scatterings. The energy transfer to the strongly-coupled optical phonons produces hot phonons, reported to exist on a femtosecond time-scale\,\cite{scops_graphite,biegert21}. Then, the photo-inverted state is established [Fig.~\ref{fig:fig1}\,(c)] through the competition between phonon-induced intraband scattering and Auger recombination~\cite{knorr2013}. The formation of the population inversion has already been thoroughly explored with tr-ARPES\,\cite{snapshots,Gierz_2015,direct_view}, which reveals its relaxation time of $\sim$100\,fs. After its decay electrons follow a Fermi-Dirac distribution at elevated temperatures [Fig.~\ref{fig:fig1}\,(d)]. The whole process of electron thermalization conceptually follows the one described for graphite in Ref.\,\citenum{bauer_18}. Subsequent hot-carrier cooling is governed by phonon-assisted scatterings on the time scale of $1-10$\,ps.


\begin{figure}[!t]
\centering
\includegraphics[width=0.5\textwidth]{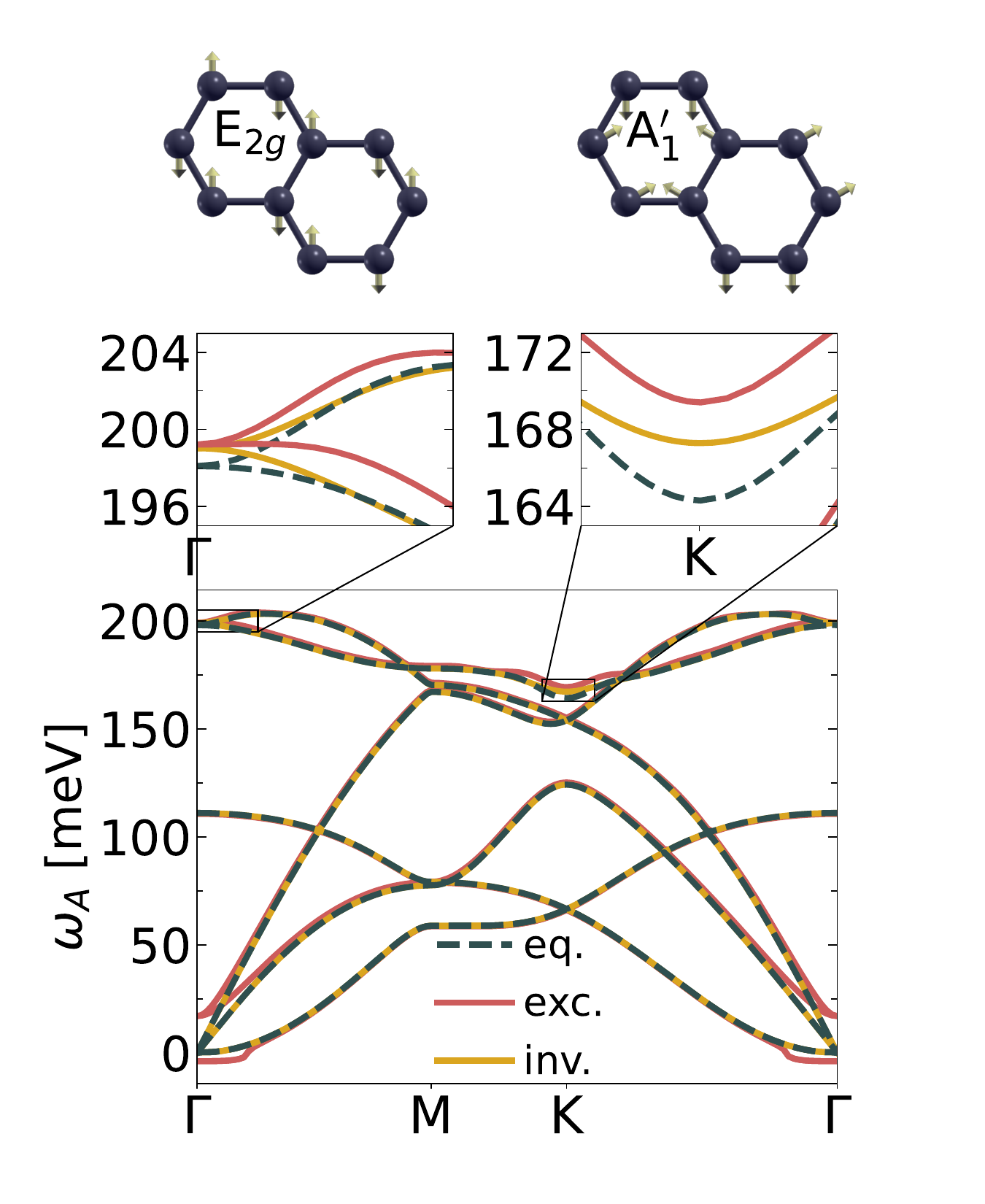}
\caption{Adiabatic DFPT phonon dispersion in the presence of a photo-excited (red line) and photo-inverted (yellow line) electron distribution in comparison with an equilibrium result (grey dashed line). Two insets are a zoom in region around high-symmetry points ($\Gamma$ and K), showing the hardening of strongly-coupled optical modes ($E_{2g}$ and $A'_{1}$) along with a schematics of the corresponding atomic motions.} 
\label{fig:fig2}
\end{figure}

\begin{figure*}[!t]
\includegraphics[width=\textwidth]{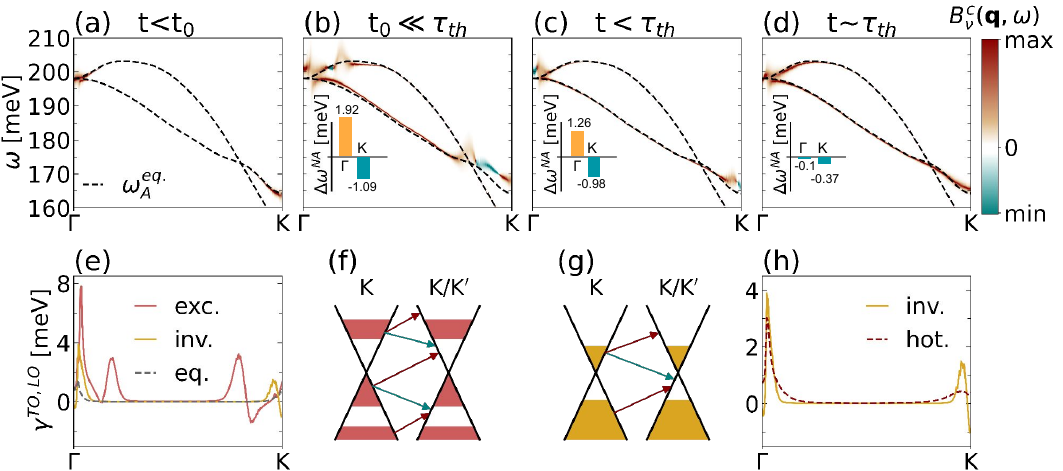}
\caption{Dynamical phonon spectral functions at different stages of electron relaxation in comparison with the adiabatic equilibrium DFPT result (grey dashed line), i.e., for: (a) Equilibrium regime, (b) far non-equilibrium following the laser excitation, (c) population inversion, (d) hot equilibrium distribution. Note the strong renormalizations occurring near the two strongly-coupled optical modes ($E_{2g}$ and $A_1'$). Also, the negative linewidth contribution to spectral function is shown in teal. The histograms in the insets of (b)-(d) show the nonadiabatic corrections to the $E_{2g}$ and $A'_{1}$ modes ($\Delta^{\rm NA}=\omega^{\rm NA}-\omega^{\rm A}$). (e) Phonon linewidth on $\Gamma$ to K path of LO/TO optical modes, due to EPC. Color-coding is the same as in Fig.~\ref{fig:fig2}. (f-g) Intra- and inter-valley (i.e., $\mathrm{K}\rightarrow\mathrm{K}$ and $\mathrm{K}\rightarrow\mathrm{K'}$) electron transitions which contribute to the phonon self-energy. The color-coded arrows reveal positive (brown) and negative (teal) contributions to the linewidth. (h) Same as in (e) but for the photo-inverted (yellow) and hot electron distributions (dark red). 
} 
\label{fig:fig3}
\end{figure*}

We implemented the nonequlibrium distributions in the calculation of electronic-structure properties and then calculated the renormalization of phonons with the \textsc{PHonon}\cite{Giannozzi2017,Baroni2001} and EPW\,\cite{Ponce2016,epw23} codes in the adiabatic and nonadiabatic approximations (see Supporting Information (SI) for more details). The resulting adiabatic phonon dispersions are shown in Fig.\,\ref{fig:fig2}. We compare the phonon dispersion for the case of photo-excited electron population as in Fig.\,\ref{fig:fig1}(b) and photo-inverted distribution Fig.\,\ref{fig:fig1}(c) with the equilibrium adiabatic case. The largest effects of the nonequilibrium electron distribution happen for the strongly-coupled optical modes $E_{2g}$ and $A'_{1}$ around $\Gamma$ and K points, respectively.  We observe phonon hardening for both phonons, and it ranges from $\simeq 1$\,meV for the $E_{2g}$ mode and $4-6$\,meV for the $A'_{1}$ mode. Our results are in a good agreement with Refs.\,\citenum{Ishioka2008,yan2009,ferrante18} and especially with Ref.~\citenum{biegert21} where the attosecond core-level spectroscopy  demonstrated that the Raman-inactive $A'_{1}$ mode is the dominating channel for dissipation of electronic coherence due to stronger coupling to electrons\,\cite{md_epc}. Note also that in our calculations the acoustic sum rule is not fully fulfilled because we inhibit the long-wavelength dipole-allowed transitions~\cite{vanLoon2021a,berges22} by filling the states above the Fermi energy. 

In Fig.~\ref{fig:fig3} we present the results of the non-equilibrium EPC calculations as obtained with cDFPT. They contain the phonon spectral function $B^{c}_{\nu}(\mathbf{q},\omega)$, which incorporates nonadiabatic renormalization effects and transverse and longitudinal optical phonon linewidth along the $\Gamma-\mathrm{K}$ path. With the term ``adiabatic'' we refer to a calculation where the phonon energy $\omega$ is omitted in the phonon self-energy calculations, and by ``nonadiabatic'' or ``dynamic'' where it is included\,\cite{Engelsberg1963,Giustino2017} (see also SI).
The first row represents four spectral functions corresponding to the four distinct electron distributions from Fig.\,\ref{fig:fig1},  together with the equilibrium adiabatic result for comparison. Equilibrium graphene linewidth contributions for the $E_{2g}$ and $A'_{1}$ modes come from the vertical interband electron scattering inside the Dirac cone ($q\simeq\Gamma$) or between the two neighboring ones ($q\simeq\mathrm{K}$). The features around $\Gamma$ and K points are symmetrical, but differ due to the disparate EPC strengths. 


The photo-excited electron distribution opens up new scattering possibilities, which are schematically shown with arrows in Fig.~\ref{fig:fig3}(f). Besides the significant modifications of the well-known nonadiabatic Kohn anomalies\,\cite{piscanec2004,lazzeri06} at the $\Gamma$ and K points coming from the non-equilibrium population, the spectral function shows additional new anomalies further away from the $\Gamma$ and K points. These photo-induced dynamical phonon anomalies come from the electron transitions away from the Dirac point, at the photo-doped regions. 
Compared to the equilibrium adiabatic dispersions, a 4\,meV renormalization is visible directly in the $\Gamma$ point for the $E_{2g}$ mode, and away from it, the highest optical branch is renormalized by 5\,meV. For the $A'_{1}$ mode, we observe a 5\,meV hardening and a 6\,meV modification at the intersection of the two optical branches. We note that these sharp transient frequency modifications are quite large and are comparable to the nonadiabatic frequency shifts of the $E_{2g}$ mode in the highly-doped graphene\,\cite{lazzeri06}. In the case of population inversion, the non-equilibrium electron distribution is condensed in the vicinity of the Dirac point, bringing the non-equilibrium spectral features closer to the $\Gamma$ and K points.  We again observe phonon renormalization for the $E_{2g}$ and $A'_{1}$ modes of about 2\,meV for both. Interestingly, for the strong non-equilibrium and population inversion we observe additional phonon hardening (softening) for $E_{2g}$ ($A_1'$) when the nonadiabatic effects are taken into account [insets of Figs.\,\ref{fig:fig3}(b) and \ref{fig:fig3}(c)]. In fact, significant increase of the nonadiabatic correction is obtained for the strong non-equilibrium, while it is reduced for the population inversion and almost diminished for the hot equilibrium case. Note that this is contrary to the conclusions drawn in Ref.\,\citenum{Ishioka2008}, where the decrease of the nonadiabaticity is suggested.

The corresponding contributions to the linewidth [Figs.\,\ref{fig:fig3}(e) and \ref{fig:fig3}(h)] show that values at the $\Gamma$ and K points are unaltered, while slightly away from these points it is significantly enhanced compared to its value at equilibrium. For highly non-equilibrium case, additional notable phonon broadening arise displaced from the high-symmetry points, at momenta where the new dynamical anomalies appear. As stated, these linewidth features stem from the electron transitions between the photo-excited block of filled states above the Dirac point, and empty block below it. 

\begin{figure*}[!t]
\includegraphics[width=\textwidth]{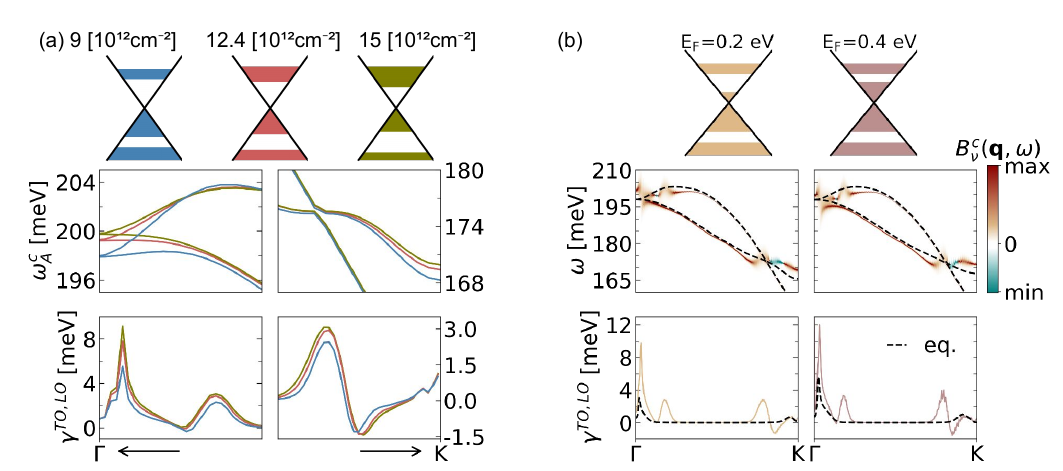}
\caption{ (a) Non-equilibrium phonon renormalization for different densities of excited electrons. In the first row, we show the corresponding adiabatic DFPT dispersions in the vicinity of $\Gamma$ and K points and observe phonon-hardening increase with photo-excited electron density. The second row contains the corresponding EPC induced linewidths. (b) Two cases of electron-doped photo-excited graphene (i.e., for $E_F=0.2$\,eV and $E_F=0.4$\,eV). In the first row, we show the corresponding dynamic spectral functions. The second row shows the linewidths for the two doping regimes. We observe the same features as in Fig.\,\ref{fig:fig3}(e) with significant increase in the linewidth close to the $\Gamma$ point. It derives from the larger electron density at the Fermi level. Again, note the negative linewidth occurrence and dynamical anomalies.  
} 
\label{fig:fig4}
\end{figure*}

A crucial thing to notice is that in the photo-excited state, electrons can scatter from the filled states at higher energies to the low-energy empty states [see Fig.~\ref{fig:fig3} (f-g), downwards pointing teal arrows], causing a negative linewidth contribution. This phonon gain, happens in the immediate vicinity of dynamical anomaly.  
For the population inversion, the phonon-gain contributions are located directly at the $\Gamma$ and K high-symmetry points.
In graphene, acoustic phonon generation was experimentally achieved~\cite{acoustic_generation} and theoretically explained~\cite{Zhao2013CerenkovEO}. 
A conceptually similar phenomenon has recently been widely explored in graphene, namely the photo-induced plasmon amplification with a high potential for the development of novel optoelectronic devices~\cite{plgain1,plgain2,plgain3,plgain4,plgain5,plgain6,plgain7,plgain8,plgain9,Duppen_2016}. Our observation of phonon gain is in agreement with the observation of the negative plasmon linewidth and negative conductivity in the photo-inverted state, and it simply means that hot phonons are emitted in the non-equilibrium regime. In particular, the results show that far non-equilibrium state supports the generation of incoherent hot phonons with momenta slightly away from the high symmetry points (i.e., phonon
displacement pattern is shifted in phase between neighbouring unit cells), while the population inversion supports coherent phonon generation of hot $E_{2g}$ and $A_1'$ (i.e., phonon
displacement pattern is repeated between neighbouring unit cells). This could explain, on the one hand, the reduction of the phonon dephasing rate of $E_{2g}$ mode as reported in Ref.\,\citenum{Ishioka2008}, and, on the other hand, the non-displacive mechanism for the generation of both $E_{2g}$ and $A_1'$ hot coherent phonons as obtained in attosecond core-level spectroscopy\,\cite{biegert21}.

As for the hot electron distribution, we calculated the number of carriers located above the Dirac point in the state of photo-inversion and found the temperature for which a Fermi-Dirac distribution produces the same number of carriers in the conduction band. We show the spectral function for the hot equilibrium electron distribution at $T=2200$\,K.  The $E_{2g}$ and $A'_{1}$ modes are slightly hardened, and here one can clearly see also the edge of the electron-hole pair excitation continuum. The linewidth resembles the one obtained for the photo-inverted population, only without the negative contributions.

Further analysis includes changing the excited carrier density, which experimentally corresponds to changing the laser fluence (Fig.~\ref{fig:fig4}). Effective carrier density is denoted in Fig.~\ref{fig:fig4}(a) above the Dirac cones, and is calculated as the summed density of photo-excited electrons and holes. As expected, we observe larger phonon stiffening in the DFPT calculation for a larger carrier density. Here we show only the adiabatic dispersions to focus solely on the increased phonon stiffening in the $\Gamma$ and K points. The $E_{2g}$ phonon hardening increases by 2\,meV as the photo-excited density increases by $6\times10^{12}$\,cm$^{-1}$. Further, we observe larger phonon linewidths deriving from the modified phase space with increasing carrier density. Directly in the $\Gamma$ point the linewidth remains at its equilibrium value, while slight differences in the position of the peaks away from the high-symmetry points are visible. 
Furthermore, since in experiments graphene is frequently placed on a substrate, we also provide results for doped graphene, specifically for Fermi levels of $E_F=200$ and 400\,meV. We compare the photo-doped spectral function with the adiabatic DFPT equilibrium calculation for the corresponding Fermi energy (dashed black lines). Again, we notice larger phonon hardening around the K point (5\,meV for both dopings) then around the $\Gamma$ point (4\,meV for $E_F=200$\,meV and 1\,meV for $E_F=400$\,meV). We observe dynamical anomalies at the same positions as they occur in the pristine photo-doped case. We notice how with increased doping, the dynamic phonon dispersion softens and the effects of photo-induced phonon hardening are less pronounced. Largest softening is observed around the dynamical anomalies. In general, for doped graphene, the intrinsic linewidths of the two highest optical modes are larger then for pristine graphene. When doped graphene is photo-excited, linewidth behaves in the same fashion as in the pristine photo-doped case, with the strong dynamical anomalies occurring in the vicinity of the high-symmetry points. 

\begin{figure*}[!t]
\includegraphics[width=\textwidth]{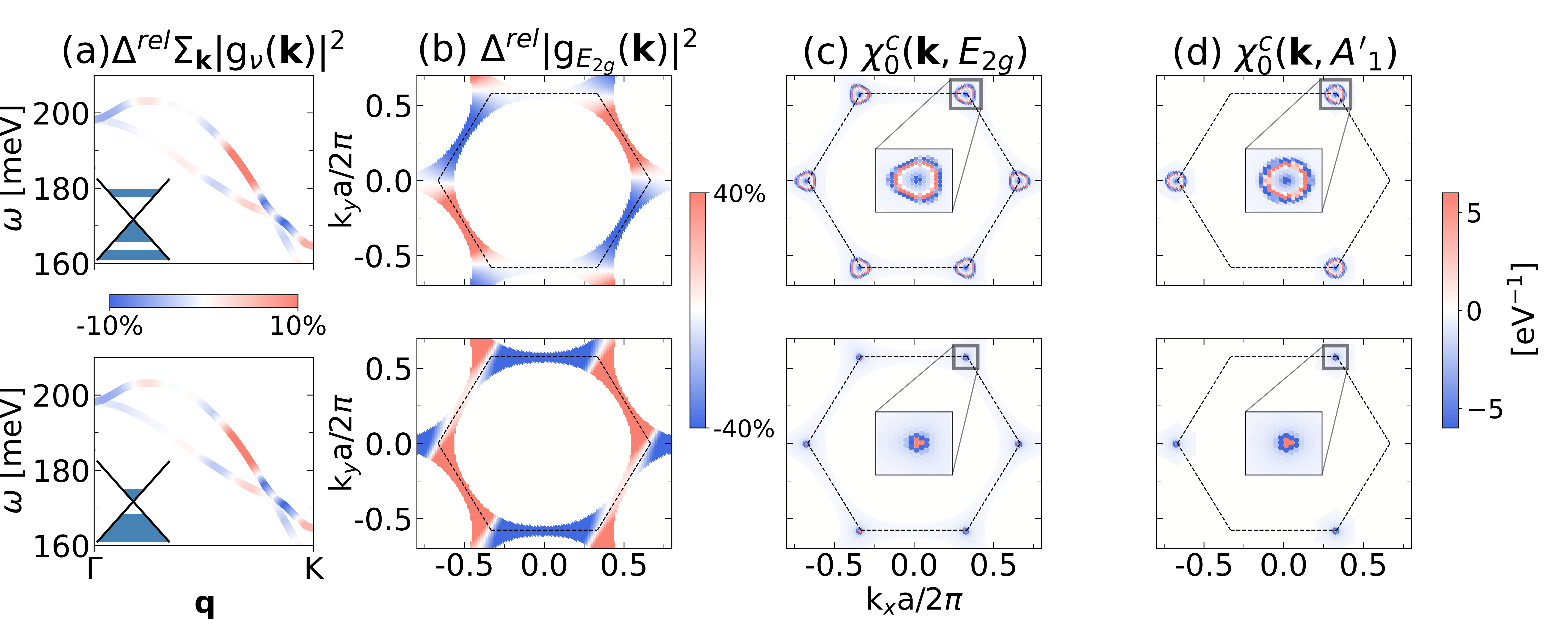}
\caption{The analysis of static phonon self-energies, revealing the electronic processes behind phonon anomalies in the case of photo-excited (first row) and photo-inverted (second-row) electron distribution. (a) Relative change in the $\mathbf{k}$-summed EPC matrix elements $\Delta^{rel}|g_{\nu}(\mathbf{k})|^2$ for the two highest optical modes and along the $\Gamma-\mathrm{K}$ q-path. (b) $\Delta^{rel}|g_{E_{2g}}(\mathbf{k})|^2$ resolved in $\mathbf{k}$ space. (c),(d) $\mathbf{k}$-resolved static susceptibility $\chi^{c}_{0}(\mathbf{k})$ for the $E_{2g}$ and $A'_{1}$ modes, respectively. The positive (negative) contributions to both $\Delta^{rel}|g_{E_{2g}}(\mathbf{k})|^2$ and $\chi^{c}_{0}(\mathbf{k})$ are shown with red (blue).
} 
\label{fig:fig5}
\end{figure*}

Finally, we present the analysis of the adiabatic phonon self-energy as calculated in cDFPT $\pi^{c}_{\nu}(\mathbf{q})$\,\cite{berges20,novko2020a} (see also SI).
Averaging out the electronic degrees of freedom deriving from the EPC matrix elements, leads to $|g_{\nu}^{nm,c}(\mathbf{k},\mathbf{q})|\rightarrow |g_{\nu}^{c}(\mathbf{q})|$ and the self-energy expression can be written as $\pi^{c}_{\nu}(\mathbf{q}) = |g^{c}_{\nu}(\mathbf{q})|^2 \chi^{c}_{0}(\mathbf{q})$, where $\chi^{c}_{0}(\mathbf{q})$ denotes the bare susceptibility function.

In this way, we can separate the non-equilibrium effects that come from the modifications in screened EPC matrix elements $|g^{c}_{\nu}(\mathbf{q})|$ and the photo-induced changes in the available phase space via $\chi^{c}_{0}(\mathbf{k})$. In Fig.~\ref{fig:fig5} we show the analysis for the conduction band, but the results for the valence band are the same due to the electron-hole symmetry. In the first two columns we show the relative difference between the EPC matrix elements in the photo-excited and photo-inverted graphene with respect to the pristine equilibrium case for the $E_{2g}$ and $A_1'$ modes $\Delta^{rel}|g_{\nu}(\mathbf{k})|^2 = (|g^c_{\nu}(\mathbf{k})|^2  - |g^{\rm eq}_{\nu}(\mathbf{k})|^2 )/{|g^{\rm eq}_{\nu}(\mathbf{k})|^2}$. In the first column, the relative differences are calculated after the electron-phonon matrix elements were summed throughout the whole Brillouin zone for a chosen $\mathbf{q}$ point on the $\Gamma - \mathrm{K}$  path and for each one of the two highest optical modes. The largest value of the relative change is only $\pm$10\% and it appears for those wavevectors $\mathbf{q}$ for which the electronic transitions are forbidden by the specific electronic structure of graphene and for which the EPC strength is weak at equilibrium (i.e., away from the $\Gamma$ and K points). More dramatic changes occur if $\Delta^{rel}|g_{E_{2g}}(\mathbf{k})|^2$ is resolved in the k$_x$-k$_y$ plane [column (b) in Fig.~\ref{fig:fig5}]. Here we explicitly show the result for $\mathbf{q}=\Gamma$ for which we found the largest values of $\Delta^{rel}|g_{E_{2g}}(\mathbf{k})|^2$ (see SI for results in additional q points). For both the photo-excited and photo-inverted electron distribution case, our calculations show that $\Delta^{rel}|g_{E_{2g}}(\mathbf{k})|^2$ reaches values of $\pm$ 40\% in certain regions of  k$_x$-k$_y$ space for the E$_{2g}$ mode. We observe a symmetrical pattern of positive and negative contributions to $\Delta^{rel}|g_{E_{2g}}(\mathbf{k})|^2$. Due to phase space restrictions, only a small region around the K points is picked out when doing a self-energy calculation, or multiplying $|g^{c}_{\nu}(\mathbf{q})|^2 $ and $\chi^{c}_{0}(\mathbf{q})$. In other words, as shown in columns (c) and (d) of Fig.~\ref{fig:fig5}, $\chi^{c}_{0}(\mathbf{k})$ is finite around the Dirac points also in a symmetric pattern. In this way, when calculating the phonon self-energy and summing over the whole Brillouin zone, the net effect of the large relative changes $\Delta^{rel}|g_{E_{2g}}(\mathbf{k})|^2$ cancels out as equal amounts of positive and negative values are picked out by the $\chi^{c}_{0}(\mathbf{k})$ factor (see Sec.\,S2 in SI for more detailed discussion).

Therefore, the effects of photo-induced changes of the electron-phonon matrix elements are small in the end, and it turns out that the phonon hardening and decrease of the dephasing rate observed in Ref.\,\citenum{Ishioka2008}, along with other non-equilibrium phenomena presented here, come dominantly from the photo-induced changes in the carrier scattering phase space $\chi^{c}_{0}(\mathbf{k})$. Therefore, optically-induced phonon blueshifts are not necessarily a sign of the suppressed coupling and it could come from the pure electronic origin, as was shown for instance in the case of photo-excited TiSe$_2$\,\cite{otto21}. This resolves the debate on whether the EPC strength is suppressed\,\cite{Ishioka2008} or enhanced\,\cite{shengmeng22} in the non-equilibrium graphene.

We elaborate this claim by thoroughly inspecting the color-coded contributions to the bare electron susceptibility $\chi^{c}_{0}(\mathbf{k})$. The features for $\mathbf{q} = \mathrm{K}$ and $\mathbf{q} = \Gamma$ are in essence the same, so the discussion is applicable to both. In the equilibrium case where the Fermi surface is almost a point, the only contribution comes from the Dirac point. Since the electrons can only scatter from the filled states below the Fermi energy to the empty states above it, the final results for $\chi^{0}(\mathbf{k})$ is negative. Focusing now on the photo-excited case (Fig.~\ref{fig:fig5}, first row), we first notice the mentioned equilibrium contribution (red dots) positioned directly in the Dirac points. Photo-excited electrons fill the states visible here as an additional triangle around each Dirac point. Each triangle consists of positive and negative susceptibility contributions. Electrons from the positive $\chi^{c}_{0}(\mathbf{k})$ portion are responsible for the negative linewidth contribution, as they contribute by scattering from the higher energy filled states to the lower energy empty ones. Due to finite temperature, this is in principle also possible within an equilibrium distribution, but those contributions are then suppressed by the much larger negative $\chi_{0}(\mathbf{k})$ contribution. The electrons from the negative $\chi^{c}_{0}(\mathbf{k})$ section make standard transitions to higher energy empty states. In the EPC calculations, we used a dynamical susceptibility term [see Eq.\,(S1) in SI], which, together with varying the wavevector $\mathbf{q}$, leads to the competition between these two contributions and, hence, to the net negative linewidth regions visible in Fig.\,\ref{fig:fig3}. The obvious difference between susceptibility contribution shape in the $\Gamma$ and K points, is the result of trigonal warping~\cite{trig_warp}, which is reversed in the neighboring K points. Setting $\mathbf{q} = \mathrm{K}$ leads to the superposition of two relatively rotated triangles, making the $A'_{1}$ susceptibility look circular. 
$\chi^{c}_{0}(\mathbf{k})$ for the photo-inverted distribution, consists of two circular contributions, as the filled/empty states are in the energy range of a linear electron distribution. Color-coding again suggests that the electrons closer to the Dirac point rather scatter to the empty states below the Fermi level, while the higher energy electrons do the opposite and contribute positively to the phonon linewidth. In this case, varying the wavevector $\mathbf{q}$ in the dynamical susceptibility calculation reduces the phase space for the electrons in the negative section, as a number of vertical transitions is restricted. It confines the phonon gain to be placed directly in the high-symmetry points.  

Note in the end that the present analysis on phonon dynamics based in cDFPT is not only restricted to the non-equilibrium induced by laser pulses, but it could be utilized to study phonon renormalization in any out-of-equilibrium conditions. For instance, impact of the static electric fields and the corresponding non-equilibrium electrons on phonons in current-carrying materials is still not well understood despite its importance for comprehending various transport properties\,\cite{sabbaghi22,mao22,mao23}.

Understanding the mechanisms behind the out-of-equilibrium EPC provides valuable insights into the fundamental physics of graphene and helps unravel the complex interplay between charge carriers and lattice vibrations.
We investigated the coupling of the high-energy optical phonon modes with the photo-excited electron distribution by means of cDFPT. We observed hardening of the well-known Kohn anomalies at the center and edge of the Brillouin zone. The latter comes from the modified phase space for electron transitions which leads to different screening effects, while the effective EPC coupling strenghts are only slightly changed. We obtained complex nonadiabatic EPC features that emerge in both the dispersion and linewidth, and mostly originate from the new scattering channels opened in non-equilibrium. For instance, sharp dynamical phonon anomalies away from the high-symmetry points and the overall increase of the phonon scattering rate have been observed. Also, we showed incoherent phonon gain at finite wavevectors irrespective of the doping level or the concentration of the photo-excited carriers, while coherent phonon generation is expected in the state of population inversion and is within the scope of typical experiments. We believe our work offers crucial information on the nature of EPC in graphene, justifies the known non-equilibrium features and sheds new light on the understanding of the underlying ultrafast vibrational relaxation mechanisms.

\begin{acknowledgement}
Useful discussions with Jan Berges and Samuel Ponc\'e are gratefully acknowledged. We acknowledge financial support from the Croatian Science Foundation (Grant no. UIP-2019-04-6869) and from the European Regional Development Fund for the ``Center of Excellence for Advanced Materials and Sensing Devices'' (Grant No. KK.01.1.1.01.0001).
\end{acknowledgement}

\begin{suppinfo}
More information on computational details and detailed analysis of the modifications in the electron-phonon matrix elements and scattering phase space due to non-equilibrium distribution\\
\end{suppinfo}

\bibliography{ms}

\providecommand{\latin}[1]{#1}
\makeatletter
\providecommand{\doi}
  {\begingroup\let\do\@makeother\dospecials
  \catcode`\{=1 \catcode`\}=2 \doi@aux}
\providecommand{\doi@aux}[1]{\endgroup\texttt{#1}}
\makeatother
\providecommand*\mcitethebibliography{\thebibliography}
\csname @ifundefined\endcsname{endmcitethebibliography}  {\let\endmcitethebibliography\endthebibliography}{}
\begin{mcitethebibliography}{75}
\providecommand*\natexlab[1]{#1}
\providecommand*\mciteSetBstSublistMode[1]{}
\providecommand*\mciteSetBstMaxWidthForm[2]{}
\providecommand*\mciteBstWouldAddEndPuncttrue
  {\def\EndOfBibitem{\unskip.}}
\providecommand*\mciteBstWouldAddEndPunctfalse
  {\let\EndOfBibitem\relax}
\providecommand*\mciteSetBstMidEndSepPunct[3]{}
\providecommand*\mciteSetBstSublistLabelBeginEnd[3]{}
\providecommand*\EndOfBibitem{}
\mciteSetBstSublistMode{f}
\mciteSetBstMaxWidthForm{subitem}{(\alph{mcitesubitemcount})}
\mciteSetBstSublistLabelBeginEnd
  {\mcitemaxwidthsubitemform\space}
  {\relax}
  {\relax}

\bibitem[Först \latin{et~al.}(2011)Först, Manzoni, Kaiser, Tomioka, Tokura, Merlin, and Cavalleri]{lattice_control}
Först,~M.; Manzoni,~C.; Kaiser,~S.; Tomioka,~Y.; Tokura,~Y.; Merlin,~R.; Cavalleri,~A. Nonlinear phononics: A new ultrafast route to lattice control. \emph{Nat. Phys.} \textbf{2011}, \emph{7}\relax
\mciteBstWouldAddEndPuncttrue
\mciteSetBstMidEndSepPunct{\mcitedefaultmidpunct}
{\mcitedefaultendpunct}{\mcitedefaultseppunct}\relax
\EndOfBibitem
\bibitem[de~la Torre \latin{et~al.}(2021)de~la Torre, Kennes, Claassen, Gerber, McIver, and Sentef]{torre21}
de~la Torre,~A.; Kennes,~D.~M.; Claassen,~M.; Gerber,~S.; McIver,~J.~W.; Sentef,~M.~A. Colloquium: Nonthermal pathways to ultrafast control in quantum materials. \emph{Rev. Mod. Phys.} \textbf{2021}, \emph{93}, 041002\relax
\mciteBstWouldAddEndPuncttrue
\mciteSetBstMidEndSepPunct{\mcitedefaultmidpunct}
{\mcitedefaultendpunct}{\mcitedefaultseppunct}\relax
\EndOfBibitem
\bibitem[Basov \latin{et~al.}(2017)Basov, Averitt, and Hsieh]{Basov2017}
Basov,~D.; Averitt,~R.; Hsieh,~D. Towards properties on demand in quantum materials. \emph{Nat. Mater.} \textbf{2017}, \emph{16}, 1077--1088\relax
\mciteBstWouldAddEndPuncttrue
\mciteSetBstMidEndSepPunct{\mcitedefaultmidpunct}
{\mcitedefaultendpunct}{\mcitedefaultseppunct}\relax
\EndOfBibitem
\bibitem[Fausti \latin{et~al.}(2011)Fausti, Tobey, Dean, Kaiser, Dienst, Hoffmann, Pyon, Takayama, Takagi, and Cavalleri]{fausti11}
Fausti,~D.; Tobey,~R.~I.; Dean,~N.; Kaiser,~S.; Dienst,~A.; Hoffmann,~M.~C.; Pyon,~S.; Takayama,~T.; Takagi,~H.; Cavalleri,~A. Light-Induced Superconductivity in a Stripe-Ordered Cuprate. \emph{Science} \textbf{2011}, \emph{331}, 189\relax
\mciteBstWouldAddEndPuncttrue
\mciteSetBstMidEndSepPunct{\mcitedefaultmidpunct}
{\mcitedefaultendpunct}{\mcitedefaultseppunct}\relax
\EndOfBibitem
\bibitem[Mitrano \latin{et~al.}(2016)Mitrano, Cantaluppi, Nicoletti, Kaiser, Perucchi, Lupi, Pietro, Pontiroli, Ricc{\`{o}}, Clark, Jaksch, and Cavalleri]{mitrano16}
Mitrano,~M.; Cantaluppi,~A.; Nicoletti,~D.; Kaiser,~S.; Perucchi,~A.; Lupi,~S.; Pietro,~P.~D.; Pontiroli,~D.; Ricc{\`{o}},~M.; Clark,~S.~R.; Jaksch,~D.; Cavalleri,~A. Possible light-induced superconductivity in K$_3$C$_{60}$ at high temperature. \emph{Nature} \textbf{2016}, \emph{530}, 461\relax
\mciteBstWouldAddEndPuncttrue
\mciteSetBstMidEndSepPunct{\mcitedefaultmidpunct}
{\mcitedefaultendpunct}{\mcitedefaultseppunct}\relax
\EndOfBibitem
\bibitem[Stojchevska \latin{et~al.}(2014)Stojchevska, Vaskivskyi, Mertelj, Kusar, Svetin, Brazovskii, and Mihailovic]{stojchevska14}
Stojchevska,~L.; Vaskivskyi,~I.; Mertelj,~T.; Kusar,~P.; Svetin,~D.; Brazovskii,~S.; Mihailovic,~D. Ultrafast Switching to a Stable Hidden Quantum State in an Electronic Crystal. \emph{Science} \textbf{2014}, \emph{344}, 177\relax
\mciteBstWouldAddEndPuncttrue
\mciteSetBstMidEndSepPunct{\mcitedefaultmidpunct}
{\mcitedefaultendpunct}{\mcitedefaultseppunct}\relax
\EndOfBibitem
\bibitem[Kogar \latin{et~al.}(2020)Kogar, Zong, Dolgirev, Shen, Straquadine, Bie, Wang, Rohwer, Tung, Yang, Li, Yang, Weathersby, Park, Kozina, Sie, Wen, Jarillo-Herrero, Fisher, Wang, and Gedik]{kogar20}
Kogar,~A. \latin{et~al.}  Light-induced charge density wave in LaTe$_3$. \emph{Nat. Phys.} \textbf{2020}, \emph{16}, 159\relax
\mciteBstWouldAddEndPuncttrue
\mciteSetBstMidEndSepPunct{\mcitedefaultmidpunct}
{\mcitedefaultendpunct}{\mcitedefaultseppunct}\relax
\EndOfBibitem
\bibitem[Nova \latin{et~al.}(2019)Nova, Disa, Fechner, and Cavalleri]{nova19}
Nova,~T.~F.; Disa,~A.~S.; Fechner,~M.; Cavalleri,~A. Metastable ferroelectricity in optically strained SrTiO$_3$. \emph{Science} \textbf{2019}, \emph{364}, 1075\relax
\mciteBstWouldAddEndPuncttrue
\mciteSetBstMidEndSepPunct{\mcitedefaultmidpunct}
{\mcitedefaultendpunct}{\mcitedefaultseppunct}\relax
\EndOfBibitem
\bibitem[Wall \latin{et~al.}(2018)Wall, Yang, Vidas, Chollet, Glownia, Kozina, Katayama, Henighan, Jiang, Miller, Reis, Boatner, Delaire, and Trigo]{VO2}
Wall,~S.~E.; Yang,~S.; Vidas,~L.; Chollet,~M.; Glownia,~J.~M.; Kozina,~M.~E.; Katayama,~T.; Henighan,~T.; Jiang,~M.~P.; Miller,~T.~A.; Reis,~D.~A.; Boatner,~L.~A.; Delaire,~O.; Trigo,~M. Ultrafast disordering of vanadium dimers in photoexcited VO$_2$. \emph{Science} \textbf{2018}, \emph{362}, 572 -- 576\relax
\mciteBstWouldAddEndPuncttrue
\mciteSetBstMidEndSepPunct{\mcitedefaultmidpunct}
{\mcitedefaultendpunct}{\mcitedefaultseppunct}\relax
\EndOfBibitem
\bibitem[Pomarico \latin{et~al.}(2017)Pomarico, Mitrano, Bromberger, Sentef, Al-Temimy, Coletti, St\"ohr, Link, Starke, Cacho, Chapman, Springate, Cavalleri, and Gierz]{gierz17}
Pomarico,~E.; Mitrano,~M.; Bromberger,~H.; Sentef,~M.~A.; Al-Temimy,~A.; Coletti,~C.; St\"ohr,~A.; Link,~S.; Starke,~U.; Cacho,~C.; Chapman,~R.; Springate,~E.; Cavalleri,~A.; Gierz,~I. Enhanced electron-phonon coupling in graphene with periodically distorted lattice. \emph{Phys. Rev. B} \textbf{2017}, \emph{95}, 024304\relax
\mciteBstWouldAddEndPuncttrue
\mciteSetBstMidEndSepPunct{\mcitedefaultmidpunct}
{\mcitedefaultendpunct}{\mcitedefaultseppunct}\relax
\EndOfBibitem
\bibitem[Hu \latin{et~al.}(2022)Hu, Zhao, Lian, Liu, Guan, and Meng]{shengmeng22}
Hu,~S.-Q.; Zhao,~H.; Lian,~C.; Liu,~X.-B.; Guan,~M.-X.; Meng,~S. Tracking photocarrier-enhanced electron-phonon coupling in nonequilibrium. \emph{npj Quantum Mater.} \textbf{2022}, \emph{7}, 14\relax
\mciteBstWouldAddEndPuncttrue
\mciteSetBstMidEndSepPunct{\mcitedefaultmidpunct}
{\mcitedefaultendpunct}{\mcitedefaultseppunct}\relax
\EndOfBibitem
\bibitem[Hase \latin{et~al.}(2002)Hase, Kitajima, Nakashima, and Mizoguchi]{hase02}
Hase,~M.; Kitajima,~M.; Nakashima,~S.-i.; Mizoguchi,~K. Dynamics of Coherent Anharmonic Phonons in Bismuth Using High Density Photoexcitation. \emph{Phys. Rev. Lett.} \textbf{2002}, \emph{88}, 067401\relax
\mciteBstWouldAddEndPuncttrue
\mciteSetBstMidEndSepPunct{\mcitedefaultmidpunct}
{\mcitedefaultendpunct}{\mcitedefaultseppunct}\relax
\EndOfBibitem
\bibitem[Thiemann \latin{et~al.}(2022)Thiemann, Sciaini, Kassen, Hagemann, Meyer~zu Heringdorf, and Horn-von Hoegen]{bismuth2}
Thiemann,~F.; Sciaini,~G.; Kassen,~A.; Hagemann,~U.; Meyer~zu Heringdorf,~F.; Horn-von Hoegen,~M. Ultrafast transport-mediated homogenization of photoexcited electrons governs the softening of the ${A}_{1g}$ phonon in bismuth. \emph{Phys. Rev. B} \textbf{2022}, \emph{106}, 014315\relax
\mciteBstWouldAddEndPuncttrue
\mciteSetBstMidEndSepPunct{\mcitedefaultmidpunct}
{\mcitedefaultendpunct}{\mcitedefaultseppunct}\relax
\EndOfBibitem
\bibitem[Ishioka \latin{et~al.}(2008)Ishioka, Hase, Kitajima, Wirtz, Rubio, and Petek]{Ishioka2008}
Ishioka,~K.; Hase,~M.; Kitajima,~M.; Wirtz,~L.; Rubio,~A.; Petek,~H. Ultrafast electron-phonon decoupling in graphite. \emph{Phys. Rev. B} \textbf{2008}, \emph{77}, 121402\relax
\mciteBstWouldAddEndPuncttrue
\mciteSetBstMidEndSepPunct{\mcitedefaultmidpunct}
{\mcitedefaultendpunct}{\mcitedefaultseppunct}\relax
\EndOfBibitem
\bibitem[Yan \latin{et~al.}(2009)Yan, Song, Mak, Chatzakis, Maultzsch, and Heinz]{yan2009}
Yan,~H.; Song,~D.; Mak,~K.~F.; Chatzakis,~I.; Maultzsch,~J.; Heinz,~T.~F. Time-resolved Raman spectroscopy of optical phonons in graphite: Phonon anharmonic coupling and anomalous stiffening. \emph{Phys. Rev. B} \textbf{2009}, \emph{80}, 121403\relax
\mciteBstWouldAddEndPuncttrue
\mciteSetBstMidEndSepPunct{\mcitedefaultmidpunct}
{\mcitedefaultendpunct}{\mcitedefaultseppunct}\relax
\EndOfBibitem
\bibitem[Krapivin \latin{et~al.}(2022)Krapivin, Gu, Hickox-Young, Teitelbaum, Huang, de~la Pe\~na, Zhu, Sirica, Lee, Prasankumar, Maznev, Nelson, Chollet, Rondinelli, Reis, and Trigo]{KTaO3}
Krapivin,~V. \latin{et~al.}  Ultrafast Suppression of the Ferroelectric Instability in ${\mathrm{KTaO}}_{3}$. \emph{Phys. Rev. Lett.} \textbf{2022}, \emph{129}, 127601\relax
\mciteBstWouldAddEndPuncttrue
\mciteSetBstMidEndSepPunct{\mcitedefaultmidpunct}
{\mcitedefaultendpunct}{\mcitedefaultseppunct}\relax
\EndOfBibitem
\bibitem[Jiang \latin{et~al.}(2016)Jiang, Trigo, Savi{\'c}, Fahy, Murray, Bray, Clark, Henighan, Kozina, Chollet, Glownia, Hoffmann, Zhu, Delaire, May, Sales, Lindenberg, Zalden, Sato, Merlin, and Reis]{Jiang2016TheOO}
Jiang,~M.~P. \latin{et~al.}  The origin of incipient ferroelectricity in lead telluride. \emph{Nat. Commun.} \textbf{2016}, \emph{7}\relax
\mciteBstWouldAddEndPuncttrue
\mciteSetBstMidEndSepPunct{\mcitedefaultmidpunct}
{\mcitedefaultendpunct}{\mcitedefaultseppunct}\relax
\EndOfBibitem
\bibitem[Otto \latin{et~al.}(2021)Otto, Pöhls, de~Cotret, Stern, Sutton, and Siwick]{otto21}
Otto,~M.~R.; Pöhls,~J.-H.; de~Cotret,~L. P.~R.; Stern,~M.~J.; Sutton,~M.; Siwick,~B.~J. Mechanisms of electron-phonon coupling unraveled in momentum and time: The case of soft phonons in TiSe$_2$. \emph{Sci. Adv.} \textbf{2021}, \emph{7}, eabf2810\relax
\mciteBstWouldAddEndPuncttrue
\mciteSetBstMidEndSepPunct{\mcitedefaultmidpunct}
{\mcitedefaultendpunct}{\mcitedefaultseppunct}\relax
\EndOfBibitem
\bibitem[Basov \latin{et~al.}(2014)Basov, Fogler, Lanzara, Wang, and Zhang]{RevModPhys.86.959}
Basov,~D.~N.; Fogler,~M.~M.; Lanzara,~A.; Wang,~F.; Zhang,~Y. Colloquium: Graphene spectroscopy. \emph{Rev. Mod. Phys.} \textbf{2014}, \emph{86}, 959--994\relax
\mciteBstWouldAddEndPuncttrue
\mciteSetBstMidEndSepPunct{\mcitedefaultmidpunct}
{\mcitedefaultendpunct}{\mcitedefaultseppunct}\relax
\EndOfBibitem
\bibitem[Rohde \latin{et~al.}(2018)Rohde, Stange, M\"uller, Behrendt, Oloff, Hanff, Albert, Hein, Rossnagel, and Bauer]{bauer_18}
Rohde,~G.; Stange,~A.; M\"uller,~A.; Behrendt,~M.; Oloff,~L.-P.; Hanff,~K.; Albert,~T.~J.; Hein,~P.; Rossnagel,~K.; Bauer,~M. Ultrafast Formation of a Fermi-Dirac Distributed Electron Gas. \emph{Phys. Rev. Lett.} \textbf{2018}, \emph{121}, 256401\relax
\mciteBstWouldAddEndPuncttrue
\mciteSetBstMidEndSepPunct{\mcitedefaultmidpunct}
{\mcitedefaultendpunct}{\mcitedefaultseppunct}\relax
\EndOfBibitem
\bibitem[Gierz \latin{et~al.}(2013)Gierz, Petersen, Mitrano, Cacho, Turcu, Springate, Stöhr, Köhler, Starke, and Cavalleri]{snapshots}
Gierz,~I.; Petersen,~J.~C.; Mitrano,~M.; Cacho,~C.; Turcu,~I. C.~E.; Springate,~E.; Stöhr,~A.; Köhler,~A.; Starke,~U.; Cavalleri,~A. Snapshots of non-equilibrium Dirac carrier distributions in graphene. \emph{Nat. Mater.} \textbf{2013}, \emph{12}, 1119--1124\relax
\mciteBstWouldAddEndPuncttrue
\mciteSetBstMidEndSepPunct{\mcitedefaultmidpunct}
{\mcitedefaultendpunct}{\mcitedefaultseppunct}\relax
\EndOfBibitem
\bibitem[Johannsen \latin{et~al.}(2013)Johannsen, Ulstrup, Cilento, Crepaldi, Zacchigna, Cacho, Turcu, Springate, Fromm, Raidel, Seyller, Parmigiani, Grioni, and Hofmann]{direct_view}
Johannsen,~J.~C.; Ulstrup,~S.; Cilento,~F.; Crepaldi,~A.; Zacchigna,~M.; Cacho,~C.; Turcu,~I. C.~E.; Springate,~E.; Fromm,~F.; Raidel,~C.; Seyller,~T.; Parmigiani,~F.; Grioni,~M.; Hofmann,~P. Direct View of Hot Carrier Dynamics in Graphene. \emph{Phys. Rev. Lett.} \textbf{2013}, \emph{111}, 027403\relax
\mciteBstWouldAddEndPuncttrue
\mciteSetBstMidEndSepPunct{\mcitedefaultmidpunct}
{\mcitedefaultendpunct}{\mcitedefaultseppunct}\relax
\EndOfBibitem
\bibitem[Na \latin{et~al.}(2019)Na, Mills, Boschini, Michiardi, Nosarzewski, Day, Razzoli, Sheyerman, Schneider, Levy, Zhdanovich, Devereaux, Kemper, Jones, and Damascelli]{na19}
Na,~M.~X.; Mills,~A.~K.; Boschini,~F.; Michiardi,~M.; Nosarzewski,~B.; Day,~R.~P.; Razzoli,~E.; Sheyerman,~A.; Schneider,~M.; Levy,~G.; Zhdanovich,~S.; Devereaux,~T.~P.; Kemper,~A.~F.; Jones,~D.~J.; Damascelli,~A. Direct determination of mode-projected electron-phonon coupling in the time domain. \emph{Science} \textbf{2019}, \emph{366}, 1231\relax
\mciteBstWouldAddEndPuncttrue
\mciteSetBstMidEndSepPunct{\mcitedefaultmidpunct}
{\mcitedefaultendpunct}{\mcitedefaultseppunct}\relax
\EndOfBibitem
\bibitem[Düvel \latin{et~al.}(2022)Düvel, Merboldt, Bange, Strauch, Stellbrink, Pierz, Schumacher, Momeni, Steil, Jansen, Steil, Novko, Mathias, and Reutzel]{reutzel22}
Düvel,~M.; Merboldt,~M.; Bange,~J.~P.; Strauch,~H.; Stellbrink,~M.; Pierz,~K.; Schumacher,~H.~W.; Momeni,~D.; Steil,~D.; Jansen,~G. S.~M.; Steil,~S.; Novko,~D.; Mathias,~S.; Reutzel,~M. Far-from-Equilibrium Electron–Phonon Interactions in Optically Excited Graphene. \emph{Nano Lett.} \textbf{2022}, \emph{22}, 4897\relax
\mciteBstWouldAddEndPuncttrue
\mciteSetBstMidEndSepPunct{\mcitedefaultmidpunct}
{\mcitedefaultendpunct}{\mcitedefaultseppunct}\relax
\EndOfBibitem
\bibitem[Xu \latin{et~al.}(1996)Xu, Cao, Miller, Mantell, Miller, and Gao]{xu96}
Xu,~S.; Cao,~J.; Miller,~C.~C.; Mantell,~D.~A.; Miller,~R. J.~D.; Gao,~Y. Energy Dependence of Electron Lifetime in Graphite Observed with Femtosecond Photoemission Spectroscopy. \emph{Phys. Rev. Lett.} \textbf{1996}, \emph{76}, 483--486\relax
\mciteBstWouldAddEndPuncttrue
\mciteSetBstMidEndSepPunct{\mcitedefaultmidpunct}
{\mcitedefaultendpunct}{\mcitedefaultseppunct}\relax
\EndOfBibitem
\bibitem[Moos \latin{et~al.}(2001)Moos, Gahl, Fasel, Wolf, and Hertel]{moos01}
Moos,~G.; Gahl,~C.; Fasel,~R.; Wolf,~M.; Hertel,~T. Anisotropy of Quasiparticle Lifetimes and the Role of Disorder in Graphite from Ultrafast Time-Resolved Photoemission Spectroscopy. \emph{Phys. Rev. Lett.} \textbf{2001}, \emph{87}, 267402\relax
\mciteBstWouldAddEndPuncttrue
\mciteSetBstMidEndSepPunct{\mcitedefaultmidpunct}
{\mcitedefaultendpunct}{\mcitedefaultseppunct}\relax
\EndOfBibitem
\bibitem[Tan \latin{et~al.}(2017)Tan, Argondizzo, Wang, Cui, and Petek]{PhysRevX.7.011004}
Tan,~S.; Argondizzo,~A.; Wang,~C.; Cui,~X.; Petek,~H. Ultrafast Multiphoton Thermionic Photoemission from Graphite. \emph{Phys. Rev. X} \textbf{2017}, \emph{7}, 011004\relax
\mciteBstWouldAddEndPuncttrue
\mciteSetBstMidEndSepPunct{\mcitedefaultmidpunct}
{\mcitedefaultendpunct}{\mcitedefaultseppunct}\relax
\EndOfBibitem
\bibitem[Kampfrath \latin{et~al.}(2005)Kampfrath, Perfetti, Schapper, Frischkorn, and Wolf]{scops_graphite}
Kampfrath,~T.; Perfetti,~L.; Schapper,~F.; Frischkorn,~C.; Wolf,~M. Strongly Coupled Optical Phonons in the Ultrafast Dynamics of the Electronic Energy and Current Relaxation in Graphite. \emph{Phys. Rev. Lett.} \textbf{2005}, \emph{95}, 187403\relax
\mciteBstWouldAddEndPuncttrue
\mciteSetBstMidEndSepPunct{\mcitedefaultmidpunct}
{\mcitedefaultendpunct}{\mcitedefaultseppunct}\relax
\EndOfBibitem
\bibitem[Breusing \latin{et~al.}(2009)Breusing, Ropers, and Elsaesser]{breusing09}
Breusing,~M.; Ropers,~C.; Elsaesser,~T. Ultrafast Carrier Dynamics in Graphite. \emph{Phys. Rev. Lett.} \textbf{2009}, \emph{102}, 086809\relax
\mciteBstWouldAddEndPuncttrue
\mciteSetBstMidEndSepPunct{\mcitedefaultmidpunct}
{\mcitedefaultendpunct}{\mcitedefaultseppunct}\relax
\EndOfBibitem
\bibitem[Pagliara \latin{et~al.}(2011)Pagliara, Galimberti, Mor, Montagnese, Ferrini, Grandi, Galinetto, and Parmigiani]{pagliara11}
Pagliara,~S.; Galimberti,~G.; Mor,~S.; Montagnese,~M.; Ferrini,~G.; Grandi,~M.~S.; Galinetto,~P.; Parmigiani,~F. Photoinduced $\pi$-$\pi^{\ast}$ Band Gap Renormalization in Graphite. \emph{J. Am. Chem. Soc.} \textbf{2011}, \emph{133}, 6318\relax
\mciteBstWouldAddEndPuncttrue
\mciteSetBstMidEndSepPunct{\mcitedefaultmidpunct}
{\mcitedefaultendpunct}{\mcitedefaultseppunct}\relax
\EndOfBibitem
\bibitem[Kampfrath \latin{et~al.}(2005)Kampfrath, Perfetti, Schapper, Frischkorn, and Wolf]{kampfrath05}
Kampfrath,~T.; Perfetti,~L.; Schapper,~F.; Frischkorn,~C.; Wolf,~M. Strongly Coupled Optical Phonons in the Ultrafast Dynamics of the Electronic Energy and Current Relaxation in Graphite. \emph{Phys. Rev. Lett.} \textbf{2005}, \emph{95}, 187403\relax
\mciteBstWouldAddEndPuncttrue
\mciteSetBstMidEndSepPunct{\mcitedefaultmidpunct}
{\mcitedefaultendpunct}{\mcitedefaultseppunct}\relax
\EndOfBibitem
\bibitem[Ferrante \latin{et~al.}(2018)Ferrante, Virga, Benfatto, Martinati, Fazio, Sassi, Fasolato, Ott, Postorino, Yoon, Cerullo, Mauri, Ferrari, and Scopigno]{ferrante18}
Ferrante,~C.; Virga,~A.; Benfatto,~L.; Martinati,~M.; Fazio,~D.; Sassi,~U.; Fasolato,~C.; Ott,~A.; Postorino,~P.; Yoon,~D.; Cerullo,~G.; Mauri,~F.; Ferrari,~A.; Scopigno,~T. Raman spectroscopy of graphene under ultrafast laser excitation. \emph{Nat. Commun.} \textbf{2018}, \emph{9}\relax
\mciteBstWouldAddEndPuncttrue
\mciteSetBstMidEndSepPunct{\mcitedefaultmidpunct}
{\mcitedefaultendpunct}{\mcitedefaultseppunct}\relax
\EndOfBibitem
\bibitem[Sidiropoulos \latin{et~al.}(2021)Sidiropoulos, Di~Palo, Rivas, Severino, Reduzzi, Nandy, Bauerhenne, Krylow, Vasileiadis, Danz, Elliott, Sharma, Dewhurst, Ropers, Joly, Garcia, Wolf, Ernstorfer, and Biegert]{biegert21}
Sidiropoulos,~T. P.~H. \latin{et~al.}  Probing the Energy Conversion Pathways between Light, Carriers, and Lattice in Real Time with Attosecond Core-Level Spectroscopy. \emph{Phys. Rev. X} \textbf{2021}, \emph{11}, 041060\relax
\mciteBstWouldAddEndPuncttrue
\mciteSetBstMidEndSepPunct{\mcitedefaultmidpunct}
{\mcitedefaultendpunct}{\mcitedefaultseppunct}\relax
\EndOfBibitem
\bibitem[Caruso \latin{et~al.}(2020)Caruso, Novko, and Draxl]{fabio_novko_20}
Caruso,~F.; Novko,~D.; Draxl,~C. Photoemission signatures of nonequilibrium carrier dynamics from first principles. \emph{Phys. Rev. B} \textbf{2020}, \emph{101}, 035128\relax
\mciteBstWouldAddEndPuncttrue
\mciteSetBstMidEndSepPunct{\mcitedefaultmidpunct}
{\mcitedefaultendpunct}{\mcitedefaultseppunct}\relax
\EndOfBibitem
\bibitem[Caruso and Novko(2022)Caruso, and Novko]{fabio_novko_22}
Caruso,~F.; Novko,~D. Ultrafast dynamics of electrons and phonons: from the two-temperature model to the time-dependent Boltzmann equation. \emph{Adv. Phys.: X} \textbf{2022}, \emph{7}, 2095925\relax
\mciteBstWouldAddEndPuncttrue
\mciteSetBstMidEndSepPunct{\mcitedefaultmidpunct}
{\mcitedefaultendpunct}{\mcitedefaultseppunct}\relax
\EndOfBibitem
\bibitem[Tong and Bernardi(2021)Tong, and Bernardi]{tong21}
Tong,~X.; Bernardi,~M. Toward precise simulations of the coupled ultrafast dynamics of electrons and atomic vibrations in materials. \emph{Phys. Rev. Res.} \textbf{2021}, \emph{3}, 023072\relax
\mciteBstWouldAddEndPuncttrue
\mciteSetBstMidEndSepPunct{\mcitedefaultmidpunct}
{\mcitedefaultendpunct}{\mcitedefaultseppunct}\relax
\EndOfBibitem
\bibitem[Murray \latin{et~al.}(2007)Murray, Fahy, Prendergast, Ogitsu, Fritz, and Reis]{Murray2007}
Murray,~E.~D.; Fahy,~S.; Prendergast,~D.; Ogitsu,~T.; Fritz,~D.~M.; Reis,~D.~A. Phonon dispersion relations and softening in photoexcited bismuth from first principles. \emph{Phys. Rev. B} \textbf{2007}, \emph{75}, 184301\relax
\mciteBstWouldAddEndPuncttrue
\mciteSetBstMidEndSepPunct{\mcitedefaultmidpunct}
{\mcitedefaultendpunct}{\mcitedefaultseppunct}\relax
\EndOfBibitem
\bibitem[Nomura and Arita(2015)Nomura, and Arita]{Nomura2015}
Nomura,~Y.; Arita,~R. Ab initio downfolding for electron-phonon-coupled systems: Constrained density-functional perturbation theory. \emph{Phys. Rev. B} \textbf{2015}, \emph{92}, 245108\relax
\mciteBstWouldAddEndPuncttrue
\mciteSetBstMidEndSepPunct{\mcitedefaultmidpunct}
{\mcitedefaultendpunct}{\mcitedefaultseppunct}\relax
\EndOfBibitem
\bibitem[Liu \latin{et~al.}(2022)Liu, Mao, Zhang, and Zhou]{hbn_cdft}
Liu,~K.; Mao,~S.; Zhang,~S.; Zhou,~J. Photoinduced Rippling of Two-Dimensional Hexagonal Nitride Monolayers. \emph{Nano Lett.} \textbf{2022}, \emph{22}, 9006--9012\relax
\mciteBstWouldAddEndPuncttrue
\mciteSetBstMidEndSepPunct{\mcitedefaultmidpunct}
{\mcitedefaultendpunct}{\mcitedefaultseppunct}\relax
\EndOfBibitem
\bibitem[Tangney and Fahy(2002)Tangney, and Fahy]{Te_A1_DFT}
Tangney,~P.; Fahy,~S. Density-functional theory approach to ultrafast laser excitation of semiconductors: Application to the ${A}_{1}$ phonon in tellurium. \emph{Phys. Rev. B} \textbf{2002}, \emph{65}, 054302\relax
\mciteBstWouldAddEndPuncttrue
\mciteSetBstMidEndSepPunct{\mcitedefaultmidpunct}
{\mcitedefaultendpunct}{\mcitedefaultseppunct}\relax
\EndOfBibitem
\bibitem[Murray \latin{et~al.}(2005)Murray, Fritz, Wahlstrand, Fahy, and Reis]{bismuth3}
Murray,~E.~D.; Fritz,~D.~M.; Wahlstrand,~J.~K.; Fahy,~S.; Reis,~D.~A. Effect of lattice anharmonicity on high-amplitude phonon dynamics in photoexcited bismuth. \emph{Phys. Rev. B} \textbf{2005}, \emph{72}, 060301\relax
\mciteBstWouldAddEndPuncttrue
\mciteSetBstMidEndSepPunct{\mcitedefaultmidpunct}
{\mcitedefaultendpunct}{\mcitedefaultseppunct}\relax
\EndOfBibitem
\bibitem[Paillard \latin{et~al.}(2019)Paillard, Torun, Wirtz, \'I\~niguez, and Bellaiche]{paillard19}
Paillard,~C.; Torun,~E.; Wirtz,~L.; \'I\~niguez,~J.; Bellaiche,~L. Photoinduced Phase Transitions in Ferroelectrics. \emph{Phys. Rev. Lett.} \textbf{2019}, \emph{123}, 087601\relax
\mciteBstWouldAddEndPuncttrue
\mciteSetBstMidEndSepPunct{\mcitedefaultmidpunct}
{\mcitedefaultendpunct}{\mcitedefaultseppunct}\relax
\EndOfBibitem
\bibitem[Marini and Calandra(2021)Marini, and Calandra]{marini_cdfpt}
Marini,~G.; Calandra,~M. Lattice dynamics of photoexcited insulators from constrained density-functional perturbation theory. \emph{Phys. Rev. B} \textbf{2021}, \emph{104}, 144103\relax
\mciteBstWouldAddEndPuncttrue
\mciteSetBstMidEndSepPunct{\mcitedefaultmidpunct}
{\mcitedefaultendpunct}{\mcitedefaultseppunct}\relax
\EndOfBibitem
\bibitem[Winzer \latin{et~al.}(2013)Winzer, Mali\ifmmode~\acute{c}\else \'{c}\fi{}, and Knorr]{knorr2013}
Winzer,~T.; Mali\ifmmode~\acute{c}\else \'{c}\fi{},~E.; Knorr,~A. Microscopic mechanism for transient population inversion and optical gain in graphene. \emph{Phys. Rev. B} \textbf{2013}, \emph{87}, 165413\relax
\mciteBstWouldAddEndPuncttrue
\mciteSetBstMidEndSepPunct{\mcitedefaultmidpunct}
{\mcitedefaultendpunct}{\mcitedefaultseppunct}\relax
\EndOfBibitem
\bibitem[Gierz \latin{et~al.}(2015)Gierz, Mitrano, Petersen, Cacho, Turcu, Springate, Stöhr, Köhler, Starke, and Cavalleri]{Gierz_2015}
Gierz,~I.; Mitrano,~M.; Petersen,~J.~C.; Cacho,~C.; Turcu,~I. C.~E.; Springate,~E.; Stöhr,~A.; Köhler,~A.; Starke,~U.; Cavalleri,~A. Population inversion in monolayer and bilayer graphene. \emph{J. Phys.: Condens. Matter} \textbf{2015}, \emph{27}, 164204\relax
\mciteBstWouldAddEndPuncttrue
\mciteSetBstMidEndSepPunct{\mcitedefaultmidpunct}
{\mcitedefaultendpunct}{\mcitedefaultseppunct}\relax
\EndOfBibitem
\bibitem[Giannozzi \latin{et~al.}(2017)Giannozzi, Andreussi, Brumme, Bunau, Nardelli, Calandra, Car, Cavazzoni, Ceresoli, Cococcioni, Colonna, Carnimeo, Corso, Gironcoli, Delugas, DiStasio, Ferretti, Floris, Fratesi, Fugallo, Gebauer, Gerstmann, Giustino, Gorni, Jia, Kawamura, Ko, Kokalj, K\"u\c{c}\"ukbenli, Lazzeri, Marsili, Marzari, Mauri, Nguyen, Nguyen, {Otero-de-la-Roza}, Paulatto, Ponc\'e, Rocca, Sabatini, Santra, Schlipf, Seitsonen, Smogunov, Timrov, Thonhauser, Umari, Vast, Wu, and Baroni]{Giannozzi2017}
Giannozzi,~P. \latin{et~al.}  Advanced capabilities for materials modelling with \textsc{Quantum ESPRESSO}. \emph{J. Phys. Condens. Matter} \textbf{2017}, \emph{29}, 465901\relax
\mciteBstWouldAddEndPuncttrue
\mciteSetBstMidEndSepPunct{\mcitedefaultmidpunct}
{\mcitedefaultendpunct}{\mcitedefaultseppunct}\relax
\EndOfBibitem
\bibitem[Baroni \latin{et~al.}(2001)Baroni, de~Gironcoli, Dal~Corso, and Giannozzi]{Baroni2001}
Baroni,~S.; de~Gironcoli,~S.; Dal~Corso,~A.; Giannozzi,~P. Phonons and related crystal properties from density-functional perturbation theory. \emph{Rev. Mod. Phys.} \textbf{2001}, \emph{73}, 515\relax
\mciteBstWouldAddEndPuncttrue
\mciteSetBstMidEndSepPunct{\mcitedefaultmidpunct}
{\mcitedefaultendpunct}{\mcitedefaultseppunct}\relax
\EndOfBibitem
\bibitem[Ponc\'e \latin{et~al.}(2016)Ponc\'e, Margine, Verdi, and Giustino]{Ponce2016}
Ponc\'e,~S.; Margine,~E.; Verdi,~C.; Giustino,~F. {EPW}: Electron--phonon coupling, transport and superconducting properties using maximally localized {Wannier} functions. \emph{Comput. Phys. Commun.} \textbf{2016}, \emph{209}, 116\relax
\mciteBstWouldAddEndPuncttrue
\mciteSetBstMidEndSepPunct{\mcitedefaultmidpunct}
{\mcitedefaultendpunct}{\mcitedefaultseppunct}\relax
\EndOfBibitem
\bibitem[Lee \latin{et~al.}(2023)Lee, Ponc{\'{e}}, Bushick, Hajinazar, Lafuente-Bartolome, Leveillee, Lian, Lihm, Macheda, Mori, Paudyal, Sio, Tiwari, Zacharias, Zhang, Bonini, Kioupakis, Margine, and Giustino]{epw23}
Lee,~H. \latin{et~al.}  Electron{\textendash}phonon physics from first principles using the {EPW} code. \emph{npj Comput. Mater.} \textbf{2023}, \emph{9}, 156\relax
\mciteBstWouldAddEndPuncttrue
\mciteSetBstMidEndSepPunct{\mcitedefaultmidpunct}
{\mcitedefaultendpunct}{\mcitedefaultseppunct}\relax
\EndOfBibitem
\bibitem[Krylow \latin{et~al.}(2020)Krylow, Hernandez, Bauerhenne, and Garcia]{md_epc}
Krylow,~S.; Hernandez,~F.~V.; Bauerhenne,~B.; Garcia,~M.~E. Ultrafast structural relaxation dynamics of laser-excited graphene: Ab initio molecular dynamics simulations including electron-phonon interactions. \emph{Phys. Rev. B} \textbf{2020}, \emph{101}, 205428\relax
\mciteBstWouldAddEndPuncttrue
\mciteSetBstMidEndSepPunct{\mcitedefaultmidpunct}
{\mcitedefaultendpunct}{\mcitedefaultseppunct}\relax
\EndOfBibitem
\bibitem[van Loon \latin{et~al.}(2021)van Loon, Berges, and Wehling]{vanLoon2021a}
van Loon,~E. G. C.~P.; Berges,~J.; Wehling,~T.~O. Downfolding approaches to electron-ion coupling: Constrained density-functional perturbation theory for molecules. \emph{Phys. Rev. B} \textbf{2021}, \emph{103}, 205103\relax
\mciteBstWouldAddEndPuncttrue
\mciteSetBstMidEndSepPunct{\mcitedefaultmidpunct}
{\mcitedefaultendpunct}{\mcitedefaultseppunct}\relax
\EndOfBibitem
\bibitem[Berges \latin{et~al.}(2022)Berges, Girotto, Wehling, Marzari, and Poncé]{berges22}
Berges,~J.; Girotto,~N.; Wehling,~T.; Marzari,~N.; Poncé,~S. Phonon self-energy corrections: To screen, or not to screen. \emph{arXiv} \textbf{2022}, \emph{2212.11806}\relax
\mciteBstWouldAddEndPuncttrue
\mciteSetBstMidEndSepPunct{\mcitedefaultmidpunct}
{\mcitedefaultendpunct}{\mcitedefaultseppunct}\relax
\EndOfBibitem
\bibitem[Engelsberg and Schrieffer(1963)Engelsberg, and Schrieffer]{Engelsberg1963}
Engelsberg,~S.; Schrieffer,~J.~R. Coupled Electron-Phonon System. \emph{Phys. Rev.} \textbf{1963}, \emph{131}, 993\relax
\mciteBstWouldAddEndPuncttrue
\mciteSetBstMidEndSepPunct{\mcitedefaultmidpunct}
{\mcitedefaultendpunct}{\mcitedefaultseppunct}\relax
\EndOfBibitem
\bibitem[Giustino(2017)]{Giustino2017}
Giustino,~F. Electron-phonon interactions from first principles. \emph{Rev. Mod. Phys.} \textbf{2017}, \emph{89}, 015003\relax
\mciteBstWouldAddEndPuncttrue
\mciteSetBstMidEndSepPunct{\mcitedefaultmidpunct}
{\mcitedefaultendpunct}{\mcitedefaultseppunct}\relax
\EndOfBibitem
\bibitem[Piscanec \latin{et~al.}(2004)Piscanec, Lazzeri, Mauri, Ferrari, and Robertson]{piscanec2004}
Piscanec,~S.; Lazzeri,~M.; Mauri,~F.; Ferrari,~A.~C.; Robertson,~J. Kohn Anomalies and Electron-Phonon Interactions in Graphite. \emph{Phys. Rev. Lett.} \textbf{2004}, \emph{93}, 185503\relax
\mciteBstWouldAddEndPuncttrue
\mciteSetBstMidEndSepPunct{\mcitedefaultmidpunct}
{\mcitedefaultendpunct}{\mcitedefaultseppunct}\relax
\EndOfBibitem
\bibitem[Lazzeri and Mauri(2006)Lazzeri, and Mauri]{lazzeri06}
Lazzeri,~M.; Mauri,~F. Nonadiabatic Kohn Anomaly in a Doped Graphene Monolayer. \emph{Phys. Rev. Lett.} \textbf{2006}, \emph{97}, 266407\relax
\mciteBstWouldAddEndPuncttrue
\mciteSetBstMidEndSepPunct{\mcitedefaultmidpunct}
{\mcitedefaultendpunct}{\mcitedefaultseppunct}\relax
\EndOfBibitem
\bibitem[Andersen \latin{et~al.}(2019)Andersen, Dwyer, Sanchez-Yamagishi, Rodriguez-Nieva, Agarwal, Watanabe, Taniguchi, Demler, Kim, Park, and Lukin]{acoustic_generation}
Andersen,~T.; Dwyer,~B.; Sanchez-Yamagishi,~J.; Rodriguez-Nieva,~J.; Agarwal,~K.; Watanabe,~K.; Taniguchi,~T.; Demler,~E.; Kim,~P.; Park,~H.; Lukin,~M. Electron-phonon instability in graphene revealed by global and local noise probes. \emph{Science} \textbf{2019}, \emph{364}, 154--157\relax
\mciteBstWouldAddEndPuncttrue
\mciteSetBstMidEndSepPunct{\mcitedefaultmidpunct}
{\mcitedefaultendpunct}{\mcitedefaultseppunct}\relax
\EndOfBibitem
\bibitem[Zhao \latin{et~al.}(2013)Zhao, Xu, and Peeters]{Zhao2013CerenkovEO}
Zhao,~C.; Xu,~W.; Peeters,~F.~M. Cerenkov emission of terahertz acoustic-phonons from graphene. \emph{Appl. Phys. Lett.} \textbf{2013}, \emph{102}, 222101\relax
\mciteBstWouldAddEndPuncttrue
\mciteSetBstMidEndSepPunct{\mcitedefaultmidpunct}
{\mcitedefaultendpunct}{\mcitedefaultseppunct}\relax
\EndOfBibitem
\bibitem[Sabbaghi \latin{et~al.}(2015)Sabbaghi, Lee, Stauber, and Kim]{plgain1}
Sabbaghi,~M.; Lee,~H.-W.; Stauber,~T.; Kim,~K.~S. Drift-induced modifications to the dynamical polarization of graphene. \emph{Phys. Rev. B} \textbf{2015}, \emph{92}, 195429\relax
\mciteBstWouldAddEndPuncttrue
\mciteSetBstMidEndSepPunct{\mcitedefaultmidpunct}
{\mcitedefaultendpunct}{\mcitedefaultseppunct}\relax
\EndOfBibitem
\bibitem[Morgado and Silveirinha(2022)Morgado, and Silveirinha]{plgain2}
Morgado,~T.~A.; Silveirinha,~M.~G. Directional dependence of the plasmonic gain and nonreciprocity in drift-current biased graphene. \emph{Nanophotonics} \textbf{2022}, \emph{11}, 4929\relax
\mciteBstWouldAddEndPuncttrue
\mciteSetBstMidEndSepPunct{\mcitedefaultmidpunct}
{\mcitedefaultendpunct}{\mcitedefaultseppunct}\relax
\EndOfBibitem
\bibitem[Ryzhii \latin{et~al.}(2007)Ryzhii, Ryzhii, and Otsuji]{plgain3}
Ryzhii,~V.; Ryzhii,~M.; Otsuji,~T. Negative dynamic conductivity of graphene with optical pumping. \emph{J. Appl. Phys.} \textbf{2007}, \emph{101}, 083114 -- 083114\relax
\mciteBstWouldAddEndPuncttrue
\mciteSetBstMidEndSepPunct{\mcitedefaultmidpunct}
{\mcitedefaultendpunct}{\mcitedefaultseppunct}\relax
\EndOfBibitem
\bibitem[Mikhailov \latin{et~al.}(2016)Mikhailov, Savostianova, and Moskalenko]{plgain4}
Mikhailov,~S.~A.; Savostianova,~N.~A.; Moskalenko,~A.~S. Negative dynamic conductivity of a current-driven array of graphene nanoribbons. \emph{Phys. Rev. B} \textbf{2016}, \emph{94}, 035439\relax
\mciteBstWouldAddEndPuncttrue
\mciteSetBstMidEndSepPunct{\mcitedefaultmidpunct}
{\mcitedefaultendpunct}{\mcitedefaultseppunct}\relax
\EndOfBibitem
\bibitem[Petersen \latin{et~al.}(2017)Petersen, Pedersen, and Javier Garc\'{\i}a~de Abajo]{plgain5}
Petersen,~R.; Pedersen,~T.~G.; Javier Garc\'{\i}a~de Abajo,~F. Nonlocal plasmonic response of doped and optically pumped graphene, ${\mathrm{MoS}}_{2}$, and black phosphorus. \emph{Phys. Rev. B} \textbf{2017}, \emph{96}, 205430\relax
\mciteBstWouldAddEndPuncttrue
\mciteSetBstMidEndSepPunct{\mcitedefaultmidpunct}
{\mcitedefaultendpunct}{\mcitedefaultseppunct}\relax
\EndOfBibitem
\bibitem[Morgado and Silveirinha(2017)Morgado, and Silveirinha]{plgain6}
Morgado,~T.~A.; Silveirinha,~M.~G. Negative Landau Damping in Bilayer Graphene. \emph{Phys. Rev. Lett.} \textbf{2017}, \emph{119}, 133901\relax
\mciteBstWouldAddEndPuncttrue
\mciteSetBstMidEndSepPunct{\mcitedefaultmidpunct}
{\mcitedefaultendpunct}{\mcitedefaultseppunct}\relax
\EndOfBibitem
\bibitem[Boubanga-Tombet \latin{et~al.}(2020)Boubanga-Tombet, Knap, Yadav, Satou, But, Popov, Gorbenko, Kachorovskii, and Otsuji]{plgain7}
Boubanga-Tombet,~S.; Knap,~W.; Yadav,~D.; Satou,~A.; But,~D.~B.; Popov,~V.~V.; Gorbenko,~I.~V.; Kachorovskii,~V.; Otsuji,~T. Room-Temperature Amplification of Terahertz Radiation by Grating-Gate Graphene Structures. \emph{Phys. Rev. X} \textbf{2020}, \emph{10}, 031004\relax
\mciteBstWouldAddEndPuncttrue
\mciteSetBstMidEndSepPunct{\mcitedefaultmidpunct}
{\mcitedefaultendpunct}{\mcitedefaultseppunct}\relax
\EndOfBibitem
\bibitem[Park \latin{et~al.}(2022)Park, Sammon, Mele, and Low]{plgain8}
Park,~S.; Sammon,~M.; Mele,~E.; Low,~T. Plasmonic gain in current biased tilted Dirac nodes. \emph{Nat. Commun.} \textbf{2022}, \emph{13}, 7667\relax
\mciteBstWouldAddEndPuncttrue
\mciteSetBstMidEndSepPunct{\mcitedefaultmidpunct}
{\mcitedefaultendpunct}{\mcitedefaultseppunct}\relax
\EndOfBibitem
\bibitem[Low \latin{et~al.}(2018)Low, Chen, and Basov]{plgain9}
Low,~T.; Chen,~P.-Y.; Basov,~D.~N. Superluminal plasmons with resonant gain in population inverted bilayer graphene. \emph{Phys. Rev. B} \textbf{2018}, \emph{98}, 041403\relax
\mciteBstWouldAddEndPuncttrue
\mciteSetBstMidEndSepPunct{\mcitedefaultmidpunct}
{\mcitedefaultendpunct}{\mcitedefaultseppunct}\relax
\EndOfBibitem
\bibitem[Duppen \latin{et~al.}(2016)Duppen, Tomadin, Grigorenko, and Polini]{Duppen_2016}
Duppen,~B.~V.; Tomadin,~A.; Grigorenko,~A.~N.; Polini,~M. Current-induced birefringent absorption and non-reciprocal plasmons in graphene. \emph{2D Mater.} \textbf{2016}, \emph{3}, 015011\relax
\mciteBstWouldAddEndPuncttrue
\mciteSetBstMidEndSepPunct{\mcitedefaultmidpunct}
{\mcitedefaultendpunct}{\mcitedefaultseppunct}\relax
\EndOfBibitem
\bibitem[Berges \latin{et~al.}(2020)Berges, van Loon, Schobert, R\"osner, and Wehling]{berges20}
Berges,~J.; van Loon,~E. G. C.~P.; Schobert,~A.; R\"osner,~M.; Wehling,~T.~O. Ab initio phonon self-energies and fluctuation diagnostics of phonon anomalies: Lattice instabilities from Dirac pseudospin physics in transition metal dichalcogenides. \emph{Phys. Rev. B} \textbf{2020}, \emph{101}, 155107\relax
\mciteBstWouldAddEndPuncttrue
\mciteSetBstMidEndSepPunct{\mcitedefaultmidpunct}
{\mcitedefaultendpunct}{\mcitedefaultseppunct}\relax
\EndOfBibitem
\bibitem[Novko(2020)]{novko2020a}
Novko,~D. Broken adiabaticity induced by Lifshitz transition in MoS$_2$ and WS$_2$ single layers. \emph{Commun. Phys.} \textbf{2020}, \emph{3}, 1\relax
\mciteBstWouldAddEndPuncttrue
\mciteSetBstMidEndSepPunct{\mcitedefaultmidpunct}
{\mcitedefaultendpunct}{\mcitedefaultseppunct}\relax
\EndOfBibitem
\bibitem[Dresselhaus(1974)]{trig_warp}
Dresselhaus,~G. Graphite Landau levels in the presence of trigonal warping. \emph{Phys. Rev. B} \textbf{1974}, \emph{10}, 3602--3609\relax
\mciteBstWouldAddEndPuncttrue
\mciteSetBstMidEndSepPunct{\mcitedefaultmidpunct}
{\mcitedefaultendpunct}{\mcitedefaultseppunct}\relax
\EndOfBibitem
\bibitem[Sabbaghi \latin{et~al.}(2022)Sabbaghi, Stauber, Lee, Gomez-Diaz, and Hanson]{sabbaghi22}
Sabbaghi,~M.; Stauber,~T.; Lee,~H.-W.; Gomez-Diaz,~J.~S.; Hanson,~G.~W. In-plane optical phonon modes of current-carrying graphene. \emph{Phys. Rev. B} \textbf{2022}, \emph{105}, 235405\relax
\mciteBstWouldAddEndPuncttrue
\mciteSetBstMidEndSepPunct{\mcitedefaultmidpunct}
{\mcitedefaultendpunct}{\mcitedefaultseppunct}\relax
\EndOfBibitem
\bibitem[Mao \latin{et~al.}(2022)Mao, Shang, and L\"u]{mao22}
Mao,~W.-H.; Shang,~M.-Y.; L\"u,~J.-T. Hot and cold phonons in electrically biased graphene. \emph{Phys. Rev. B} \textbf{2022}, \emph{106}, 125406\relax
\mciteBstWouldAddEndPuncttrue
\mciteSetBstMidEndSepPunct{\mcitedefaultmidpunct}
{\mcitedefaultendpunct}{\mcitedefaultseppunct}\relax
\EndOfBibitem
\bibitem[Mao \latin{et~al.}(2023)Mao, Shang, and L\"u]{mao23}
Mao,~W.-H.; Shang,~M.-Y.; L\"u,~J.-T. Current-driven collective dynamics of non-Hermitian edge vibrations in armchair graphene nanoribbons. \emph{Phys. Rev. B} \textbf{2023}, \emph{107}, 085419\relax
\mciteBstWouldAddEndPuncttrue
\mciteSetBstMidEndSepPunct{\mcitedefaultmidpunct}
{\mcitedefaultendpunct}{\mcitedefaultseppunct}\relax
\EndOfBibitem
\end{mcitethebibliography}

\end{document}